\documentclass[11pt]{amsart}
\usepackage{geometry}                
\usepackage{graphicx}
\usepackage{amssymb}
\usepackage{epstopdf}
\usepackage{listings}
\usepackage{xcolor}
\usepackage{hyperref}
\usepackage{courier}

\definecolor{darkgreen}{RGB}{0,102,0}
\definecolor{darkblue}{RGB}{0,0,102}
\definecolor{darkgray}{RGB}{96,96,96}
\definecolor{teal}{RGB}{0,139,139}
\definecolor{soothingred}{RGB}{255,0,107}

\usepackage[T1]{fontenc}

\title{Supervised Text Classification \\ using \\ Text Search}

\author{Nabarun Mondal, Mrunal Lohia}

\begin{document}
\maketitle

\begin{abstract}

Supervised text classification is a classical and active area of ML research \cite{cb}. 
In large enterprise, solutions to this problem has significant importance. 
This is specifically true in ticketing systems where 
prediction of the type and subtype of tickets given new incoming ticket text 
to find out optimal routing is a multi billion dollar industry \cite{ssa}. 

 In this paper authors describe a class of industrial standard 
algorithms which can accurately ( 86\% and above )  predict classification
of any text given prior labelled text data - by novel use of any text search engine.

  These algorithms were used to automate routing of issue tickets to the appropriate team. 
This class of algorithms has far reaching consequences for a wide variety of industrial applications, 
IT support, RPA script triggering, even legal domain 
where massive set of pre labelled data are already available.

\end{abstract}

\begin{section}{Text Classification Problems for Enterprises}

\begin{subsection}{Text in Enterprise}
Text is the fundamental mode of information exchange for any enterprise. 
Mode of the medium of text changes from Enterprise to Enterprise. 
For example older Enterprise work with e-mails, 
while modern enterprises work with chat environments like slack \& SMSes.

One large section of Enterprise text data is stored in it's ticketing (issue/incident management) systems.
In this scenario many issues ( tickets ) are created verbatim via humans by 
copying the text from emails (manual / automated) and almost always 
manually classifying the resulting tickets into class and subclass ( type and subtype ) 
into the ticketing system.

For large enterprises this manual effort pose significant problem. 
Imagine a reasonably sized enterprise where 10,000  people 
are active - and close to 10\% of them are opening tickets per hour. 
This implies, close to 100 tickets will be created per hour. 
This results in keeping a dedicated team of people to comprehend meaning 
of the texts to classify the ticket manually, so that the tickets can be 
routed into appropriate category/sub category.

Given there are close to millions of existing tickets contains text 
and the ticket was classified already ( category / sub category ) 
a plethora of pre labelled data is already available. 
At the same time, text search is implicit in any ticketing system, 
and if not, any standard search component like Elastisearch can be used to index the past text, 
``free text searching'' is a solved problem in any Enterprise.

Hence, it was essential to find out if there was a way to 
piggyback on free text search  algorithms to classify text, 
given there are already pre-existing labelled texts ( in form of tickets ).

\end{subsection}

\begin{subsection}{Formal Problem}
The formal problem can be then surmised as such:
There is a list of pre-existing \emph{independent labels} : $L = \{ \lambda_j \}$.

There is a past labelled set of tickets already: $D = \{  d_i \}$ where each $d_i :=< t_i,  l_i \subseteq L>$
where $t_i$ is the text of the ticket $d_i$ , while $l_i$ is a subset of labels the text $t_i$ tagged with.

Suppose:
$t_I$ is the input text, newer text whose labels we need to predict.
$l_p$ will be the predicted labels.
Given $D$ as training data, objective is to find a 
learning (Classification) function $\mathcal{C}$, such that, 
given past historical data and input text such that: 

$$
l_p := \mathcal{C} (t_I, D ) 
$$
\end{subsection}

As we have received inspiration for the algorithm class from ``Exit Poll''s 
We call them {\bf Exit Poll Algorithms} Family.

\end{section}

\begin{section}{Exit Poll Algorithms}

\begin{subsection}{Why Exit Poll?}

Inspiration of all algorithms described in the current paper comes from ideas of polling in democracy.

Imagine selection of the winner (elected)  from the prediction from exit poll. 
Not every individual can be sampled - and we can not trust what they are saying.
But given large enough sample, we can expect that things will cancel out and eventually the sample 
should be a good representation of the underlying population.

Can we now predict what sort of population will elect what sort of candidate?
This idea is the foundation for the exit poll set of algorithms.
It is obvious that individual poll points are similar with tickets, 
while the labels assigned to the tickets similar to candidate selection.

Hence, we ask the following question, given the input text, 
let the underlying search engine search the similar texts, 
and if the search engine is unbiased enough - 
the list of search results returned can be used to make a decision 
on what category the input text should be put in - based on the resultant items labels.

Next we showcase that how search algorithm can be used as a sampler from text.
\end{subsection}

\begin{subsection}{Search Algorithm As Sampler \& Population Identifier}

A search algorithm works as follows. Given a search text $t_I$ and training data $D$, 
it lists out \emph{similar} texts \cite{cb}.

That is, there is a pseudometric $\delta$ defined such that $\delta(t_i, t_j) \in [0,1]$ 
defines the pseudometric distance \cite{als} \cite{cv} between two texts $t_i, t_j$. 
Search algorithm $\mathcal{S}$ is a function that takes $t_I$ as input and produce a set of $d_x$ such that:
$$
\{ d_x \} := \mathcal{S}(t_I, D) 
$$
where 
$$
d_x := < t_x, \delta_x , l_x >
$$
with $\delta_x \in [0,1]$ is the match score for the data item $d_x$ against $t_I$.
Hence it makes sense to use a cut-off $\delta_c$ such that:
$$
\{ d_x \} := \forall d_x \; \delta_x < \delta_c
$$ 
which essentially ignore the more dissimilar ones.

Thus we state the working principle for our algorithm: 
{\bf Similar texts should be classified similarly.}

Thus, if a text $t_I$ has \emph{neighbours} \cite{bgrs}
$\mathcal{N}(t_I) = \{ d_x : d_x \in D \; ; \; \delta_x < \delta_c \}$ then it is expected that 
the text label must be \emph{similar} to what the labels are in $\mathcal{N}(t_I)$.

\end{subsection}

\begin{subsection}{Axiom Of Mixing}

All the algorithms assume a form of generative model, 
coming from the idea of mixing of different labels while the text gets generated. 

Notice the setting of the prediction problem,
that there are pre-existing labels $L = \{ \lambda_j \}$.
Clearly these labels are underlying root cause of the problems the tickets are depicting.
Suppose a text $t$ has labels $l = \{ \lambda_k \}$.
Suppose $\mathcal{G}( \lambda_x )$ generates stochastic text for the label $\lambda_x$. 

Then, we can imagine a generative model which generates the text $t$ from $l$ as follows:

$$
t =  \mathcal{M}( \mathcal{G}(\lambda_1) , ..., \mathcal{G}(\lambda_n) ) 
$$

where $\mathcal{M}$ is some mixing operator on texts.
Hence, we can assume that the text $t$ gets generated by Mixing the different texts generated 
individually by the underlying text generators $\mathcal{G}(\lambda_k)$. 

\end{subsection}

\begin{subsection}{All Possible Labels}

A reasonable $\mathcal{S}$ isolates all possible $d_x$ having $t_x$ \emph{similar to the input text} $t_I$,
which can be used to generate the set of plausible labels $\Lambda_p$ 
can be given by the following algorithm.

{\bf Inputs: }
\begin{enumerate}
\item{ Past Labelled Data: $D$ }
\item{ Text to classify: $t_I$ }
\item{ Search Algorithm: $\mathcal{S}(t_I,D)$ }
\end{enumerate}

{\bf Steps: }
\begin{enumerate}
\item{ Run $\mathcal{S}$ on $(t_I,D)$ :
$$ 
\{ d_x \} := \mathcal{S}(t_I, D)
$$ 
Let $l(d_x) = l_x$, the label set of the data $d_x$. 
}
\item{ All plausible labels are given by the union:

$$
\Lambda_p = \{ \lambda_x \} := \bigcup_x l_x
$$
}
\end{enumerate}

$\Lambda_p$ being the set of all possible labels which one can associate 
with the input text $t_I$ under the search algorithm $\mathcal{S}$ 
with the cut-off $\delta_c$.
\end{subsection}
\end{section}

\begin{section}{Likelihood : Prediction of top-k Labels}

\begin{subsection}{Labels with Maximal Chances}

Previous section describes the naive approach to isolate the predicted 
labels $\Lambda_p $, however, does not associate any probability distribution ( or likelihood ) 
with it. This pose a problem. Naturally, given 
some text, some labels are more likely than others - while others are less likely.

The search algorithm $\mathcal{S}$ , by definition, will do a reasonable job of eliminating 
labels which are almost unlikely. However, comparison still needs to be made between
likely labels to restrict the max number of possible labels to a desired minimum. 
This section solves that problem.   

\end{subsection}

\begin{subsection}{Naive Majority}
One extremely simplistic approach is Naive majority approach.
In this approach a counter is kept for each labels $ \lambda_x \in \Lambda_p$.
This counter is generated from the $\mathcal{S}(t_I, D)$ results where for 
each $d_x$ we find the $l_x$ and increment counter for each label.

Finally, the $\lambda_M$ having maximum frequency in the \emph{sample} wins \cite{res}.
In case there are ties, one of the competing labels is randomly selected. 

In case of top-k needs to be selected, sorting the labels by frequency of occurence
and selecting top-k does the trick.

This strategy, albeit extremely simple, works very well in a large sample where all
labels are not almost uniformly present.
\end{subsection}

\begin{subsection}{Weighted Quorum}
Naive Majority does not consider the pseudometric distance between the text and the results at all.
Weighted Quorum handles that problem by introducing a weight while counting the frequency.

Given the results of the $\mathcal{S}$ produces normalised scores ( pseudometric ) as follows:

$$
\{ d_x \}  = \mathcal{S}(t_I, D) \;  ; \; d_x := < t_x, l_x, \delta_x > \; ; \; \delta_x \in [0,1]
$$

where $\delta_x$ is the pseudometric, $\delta_x = d_p(t_x, t_I)$.   
Thus, the closer $\delta_x$ is to $0$, 
the more confidence we have in that $t_I$ and $t_x$ are very similar. 
Suppose a label $\lambda_k$ occurred with $\{ \delta_{jk} \} $ scores.   
A weighted metric $\mathcal{W}$ for each $\lambda_k$ can be created as follows:

$$
\mathcal{W}(\lambda_k) := \sum_j ( 1 - \delta_{jk} )
$$

Now, we sort descending the $\mathcal{W}(\lambda_k)$ and find the $\lambda_k$ with maximal score \cite{gtvl}.
In case of a tie, we randomly select any. Same can be extended for finding top-k.

\end{subsection}

\begin{subsection}{Unbiased Quorum}
Previous algorithm does not take into consideration that there might be an 
inherent bias in the training data itself. That is, quite possible one particular label 
might be overpopulated and hence over represented, while some might be underpopulated, 
hence under represented.

This can be a problem if the search algorithm itself is not properly tuned to generate 
the right score, hence the weighted quorum won't be able to solve it.

This can be be solved by {\bf boosting} \cite{zjs}. 

Suppose the algorithm $\mathcal{S}$ does only random sampling. 
Then, the probability that a label would be selected at random is given by the frequency 
of occurrence of the label $\lambda_k$ in the data set : $f_k$, and if the data size is
given by $N$, then the naive probability of occurrence is given by $p_k = f_k/N$.
We define boosted weighted quorum as follows:

$$
\mathcal{W_B}(\lambda_k) := \frac{1}{p_k} \sum_j ( 1 - \delta_{jk} )
$$

which ``boosts'' against the population bias.
Now, one can use this $\mathcal{W_B}$ to find out the most likely labels for the text.

\end{subsection}

\end{section}

\begin{section}{Results \& Summary}

\begin{subsection}{Production Results \& Experiments}
We have used the basic Naive Version of this algorithm for $0.2$ million tickets
with an accuracy of 95\%. While applying the same for much smaller data, 
it became apparent that the weighting is needed, and we have found that even for much smaller data set, 
with intrinsic population bias, the weighted quorum gives close to 87\% accuracy.
    
Unbiased quorum while does not improve the accuracy, 
but stabilises the algorithm against biases. 

\end{subsection}
   
\begin{subsection}{Why This Algorithm Matters}
Clearly, there are better class of classification algorithm possible.
Unfortunately none of the algorithms does multi class segregation well, 
many fails due to the curse of the dimensionality \cite{rm}. 

In contrast to those novel algorithms, this search based algorithm 
without any training requirement or any special hardware requirement
able to furnish close to 90\% accuracy on many cases. Worst case noted 
accuracy was 78\%, due to bad data health.

A worst case accuracy of 80\% is more than good enough for 
almost all practical enterprise class applications.
While it is well understood that this algorithm does not improve on 
accuracies of existing binary classifiers - but it works very well 
for multi class problems.

Moreover, given it uses an underlying free text search 
as a component - and is extremely simple - it can have widespread application 
as a convenience method.

\begin{subsection}{Further Research}
As far as the authors are concerned no such class of algorithm based on text search 
is currently in use or known in literature.

This paper was not intended to deep dive into the mechanism
of how this algorithms works. We have established some intuitive notion 
of how things are working - but did not do enough to come up with a provable formal model.
Those research are yet to be done.

\end{subsection}

\end{subsection}

\end{section}

\end{document}